# Diverse Albedos of Small Trans-Neptunian Objects

W.M. Grundy
Lowell Observatory, 1400 W. Mars Hill Rd., Flagstaff AZ 86001

K.S. Noll and D.C. Stephens[1]
Space Telescope Science Institute, 3700 San Martin Dr., Baltimore MD 21218

[1.] Now at Johns Hopkins University, Baltimore MD 21218

Primary contact: Will Grundy
E-mail: W.Grundy@lowell.edu
Telephone: 928-774-3358
Fax: 928-774-6296

Running head: diverse TNO albedos
Manuscript pages: 27
Figures: 1
Tables: 2

## ABSTRACT

Discovery of trans-neptunian object (TNO) satellites and determination of their orbits has recently enabled estimation of the size and albedo of several small TNOs, extending the size range of objects having known size and albedo down into the sub-100 km range. In this paper we compute albedo and size estimates or limits for 20 TNOs, using a consistent method for all binary objects and a consistent method for all objects having reported thermal fluxes. As is true for larger TNOs, the small objects show a remarkable diversity of albedos. Although the sample is limited, there do not yet appear to be any trends relating albedo to other observable properties or to dynamical class, with the possible exception of inclination. The observed albedo diversity of TNOs has important implications for computing the size-frequency distribution, the mass, and other global properties of the Kuiper belt derived from observations of objects' apparent magnitudes and may also point the way toward an improved compositional taxonomy based on albedo in addition to color.

## 1. Introduction

Considerable progress has been made in understanding the dynamical structure of the trans-neptunian solar system, but the physical characteristics of the small bodies that inhabit that region remain poorly determined. Measurements of colors, absolute magnitudes, and lightcurves have accumulated for a number of objects, and near infrared spectra have been reported for some of the brightest ones, but knowledge of fundamental properties such as size, mass, albedo, and density remain woefully sparse. Determining these properties for a representative sample of TNOs is crucial for estimating the total mass of material in the trans-neptunian region, for relating magnitude-frequency distributions to size- and mass-frequency distributions, for elucidating patterns of compositional taxonomy, for quantitative interpretation of infrared spectra, and for constraining the internal compositions of these objects. All of these objectives have important implications for physical and chemical conditions in the outer proto-planetary nebula and for the accretion of solid objects in the outer solar system.

Data constraining TNO sizes, albedos, and colors are compiled in this paper in an effort to search for possible patterns related to TNO origins or subsequent surface processing. We use a consistent set of models to derive size and albedo constraints from thermal emission observations and from binary orbits reported by various authors, thereby expanding the range of diameters included in the sample to more than an order of magnitude.

Although they likely do derive from the TNO population, Centaurs (by which we mean non-resonant objects which cross the orbits of major planets) and comet nuclei were excluded from this study because we lack knowledge about what part of the trans-neptunian system they originated from. Comet nuclei also have systematically different colors from TNOs (e.g., Jewitt 2002), inferred to result from volatile loss processes which may also affect Centaurs.

## 2. Size and albedo

Size and albedo are fundamental properties of bodies which constrain composition and internal structure. If systematic patterns were found to exist among these properties, they could reveal much about the accretion of TNOs in the proto-planetary nebula and about the subsequent processing of their surfaces. Different types of TNOs could have formed at different rates or in different source regions resulting in different bulk compositions. Objects in different types of orbits could experience different cratering rates or radiolytic fluxes (e.g., Stern 2002a; Cooper *et* *al.* 2003). Smaller objects may see their surfaces eroded by impacts while larger objects retain impact ejecta (e.g., Stern 1995; Durda and Stern 2000). Larger objects could also have their surface compositions altered by thermal processes such as volatile transport or differentiation (e.g., De Sanctis *et al.* 2001). It has been postulated that TNO colors reflect surface age; that their surfaces darken and redden over time via radiolysis and photolysis, in contrast with fresher surfaces which feature recently exposed, bright ices, and so appear more gray (e.g., Luu and Jewitt 1996; Gil-Hutton 2002; Peixinho *et al.* 2003; Strazzulla *et al.* 2003; Thébault and Doressoundiram 2003; Delsanti *et al.* 2004). Alternatively, gray surfaces could be so highly radiation damaged and devolatilized that they are extremely dark, with an almost pure carbon composition (e.g., Johnson *et al.* 1987; Moroz *et al.* 2003). Impacts by micrometeoroids could also play an important role in churning up gray or red subsurface materials (Cooper *et al.* 2003). Comparing albedo with other parameters for a large sample of TNOs could potentially provide useful information for these types of inquiries.

Size and albedo are, unfortunately, extremely difficult to measure for TNOs. To date, data enabling estimates or limits for 20 TNOs has been obtained from a variety of sources and methods. Despite potential risks in comparing data having different systematic uncertainties, the growing body of objects having constrained size and albedo offers an opportunity to look for potential correlations between albedo and orbital parameters, size, or color. In the following subsections we describe the sources of size and albedo constraints used in this paper along with their characteristic uncertainties.

### *2.1 Thermal observations*

Simultaneously observing thermal emission and reflected sunlight is a well-established technique which has been widely used to estimate asteroid size and albedo (e.g., Lebofsky and Spencer 1989). A thermal model (subject to various assumptions) can be inverted to solve for a size and an albedo which are consistent with observed emitted and reflected fluxes. Principal sources of uncertainty in TNO radiometric diameters include lack of knowledge of pole orientation and thermal inertia.

Thermal emission from trans-neptunian objects is extremely difficult to detect, owing to their great distance, small size, and low surface temperatures. Blackbody radiation from TNOs typically peaks in the vicinity of 100 μm, a wavelength where ground-based observations are precluded by the opacity of the terrestrial atmosphere. Nevertheless, thermal flux measurements have now been reported for 4 TNOs, and upper limits have been reported for 9 more. Thomas *et al.* (2000) reported a 90 μm thermal measurement from the ISO spacecraft for (15789) 1993 SC along with a tentative detection of (15874) 1996 $TL_{66}$ which we take to be an upper limit. Additionally, various ground-based observers have detected the Rayleigh-Jeans tail of thermal emission from TNOs at sub-mm wavelengths, including (20000) Varuna, observed by Jewitt *et al.* (2001) and by Lellouch *et al.* (2002). Margot *et al.* (2002) have reported 1.2 mm observations of (55565) 2002 $AW_{197}$. Altenhoff *et al.* (2004) reported 1.2 mm observations of (47171) 1999 $TC_{36}$ and upper limits for (28978) Ixion, (24835) 1995 $SM_{55}$, (19308) 1996 $TO_{66}$, (19521) Chaos, (38628) Huya, and (84522) 2002 $TC_{302}$. Ortiz *et al.* (2004) have reported an upper limit for (55636) 2002 $TX_{300}$. Brown *et al.* (2004) have reported an upper limit for (90377) Sedna. Many additional thermal observations of TNOs can be expected over the next few years from the Spitzer Space Telescope (SST), although SST targets tend to be biased towards the larger and nearer objects which are more readily detected at thermal infrared wavelengths. An additional advantage of Spitzer observations is its ability to observe more than one thermal wavelength, which can provide the additional constraint needed to overcome many of the uncertainties associated with thermal properties and pole orientation (e.g., Lebofsky and Spencer 1989; Stansberry *et al.* 2004).

For this paper, we fitted a consistent suite of thermal models to reported thermal fluxes in conjunction with published visual photometry (from sources detailed in Section 3). Upper limits on diameters (and lower limits on albedos) were derived by using an infinite thermal inertia ("fast rotator") model without beaming, in equator-on orientation, while the opposite limits were derived from a zero thermal inertia ("slow-rotator") model augmented with a beaming parameter of 0.8. These two cases conservatively encompass the gamut of plausible thermal models. The most probable thermal model was taken to be a fast-rotator tilted 30 degrees away from equatorial orientation.

### 2.2 Direct imaging

The diameter of the large Classical KBO (50000) Quaoar has been determined via direct observation by Brown and Trujillo (2004), using the Hubble Space Telescope's Advanced Camera for Surveys High Resolution Camera (HST/ACS HRC). An upper limit to the diameter of (90377) Sedna can be set from HRC observations that failed to resolve that object. At the time of the HST observation, Sedna was at a geocentric distance of 90.375 AU. At this distance, the ACS/HRC y-axis pixel scale of 0.0247 arcsec pixel$^{-1}$ corresponds to a linear distance of 1,620 km. The corresponding x-axis values are 0.02855 arcsec pixel$^{-1}$ and 1870 km. An object of this size or larger would show extension perpendicular to the motion vector (the observations did not track Sedna). An approximate upper limit to the diameter can be set at $d < 1800$ km from the ACS/HRC data, similar to the upper limit diameter inferred from the reported non-detection by SST (Brown *et al.* 2004). The principal source of uncertainty in direct imaging diameter measurements comes from lack of knowledge of a target's center-to-limb brightness profile.

### *2.3 Binaries*

An exciting new source of TNO sizes and albedos is astrometric observations of TNOs with natural satellites (e.g., Noll 2003). Determining a satellite's orbital period and semimajor axis yields the system mass. Total system volume can then be computed for an assumed density. For this paper we used densities of 0.5 and 2 g $cm^{-3}$ to bracket the plausible range of system densities. We then used relative visual photometry between primary and secondary to compute their relative sizes, assuming both have the same albedo, and used published absolute photometry to constrain that albedo. Because the majority of the surface area in a binary is presented by the primary body, its diameter and albedo are better constrained than are those of the secondary. For this reason, sizes and albedos of binary secondaries are not considered in this paper.

Binaries are especially valuable in providing data for objects much too small to detect with current thermal infrared or direct imaging technologies. The first binary orbit was reported by Veillet *et al.* (2002) for 1998 $WW_{31}$. Since then, orbits have been determined for (66652) 1999 $RZ_{253}$ (Noll *et al.* 2004a), (58534) 1997 $CQ_{29}$ (Margot *et al.* 2004; Noll *et al.* 2004b), (88611) 2001 $QT_{297}$ (Osip *et al.* 2003), (26308) 1998 $SM_{165}$, (47171) 1999 $TC_{36}$, and 2001 $QC_{298}$ (Margot *et al.* 2004). Future observations of thermal emission from binaries, mutual events, or stellar occultations would greatly help in constraining TNO densities, by providing independent determinations of their sizes (e.g., Noll 2003; Elliot and Kern 2003).

## 3. Data processing

By using a consistent set of thermal models to re-compute sizes and albedos from reported thermal and visual fluxes and a consistent set of densities to estimate sizes and albedos from system masses and visual photometry we attempted provide as consistent as possible a basis for comparison. We also used photometry from the literature to compute the spectral slope parameter *s*, defined as the percent increase in reflectance per 100 nm of wavelength relative to the V central wavelength 550 nm, for wavelengths ranging from V to I (e.g., Boehnhardt *et al.* 2001). We computed *s* from all available photometry in that wavelength range, combined according to reported uncertainties. Values of *s* for TNOs range from less than 5 for gray objects to greater than 10 for red objects. Table 1 lists dynamical class, spectral slope *s*, method of constraining size and albedo, diameter *d*, and R albedo for the 20 TNOs having published size and albedo constraints, as well as the planet Pluto and its satellite Charon. Data sources for specific objects are detailed in Table 2 and for objects requiring non-standard treatment, in the remainder of this section.

Note to Editor: Tables 1 and 2 should go somewhere near here.

For (15874) 1996 $TL_{66}$, we took the radiometric flux reported by Thomas *et al.* (2000) to be an upper limit because nothing was seen at the object's ephemeris location.

For (20000) Varuna, we used radiometric fluxes from Jewitt *et al.* (2001) and Lellouch *et al.* (2002) separately, and then took the resulting albedo and diameter limits to encompass the range of possible values.

For the binary (47171) 1999 $TC_{36}$, we used the Margot *et al.* (2004) mass and differential photometry. A radiometric flux measurement by Altenhoff *et al.* (2004) offers an independent measure of the projected surface area of (47171) 1999 $TC_{36}$ which can be combined with the system mass to compute an apparently absurd density of $0.15\ {}^{+0.17}_{-0.05}$ g cm$^{-3}$ for the primary. However, the Altenhoff *et al.* (2004) size has been called into question by Stansberry *et al.* (2004), who find a size consistent with our size estimates from the system mass (Stansberry personal communication 2004).

For (50000) Quaoar, we used the direct imaging diameter of Brown and Trujillo (2004).

For (55565) 2002 $AW_{197}$, Margot *et al.* (2002) reported a radiometric diameter and R albedo but did not report the observed thermal flux, so we could not apply our standard thermal models to this object, and instead used the albedo and size reported by Margot *et al.* Recent results from Spitzer Space Telescope reported by Stansberry *et al.* (2004) suggest that this object has a somewhat smaller diameter and higher albedo than found by Margot *et al.* (2002).

For (84522) 2002 $TC_{302}$, we were unable to find photometric colors to compute a spectral slope. The only available photometry for this object was reported in MPECs 2002-V26 and 2002-X65.

For (90377) Sedna we used the Brown *et al.* (2004) radiometric limits and photometry, which are consistent with direct imaging limits based on HST/ACS observations. Brown *et al.* did not report their thermal flux limits, so we were unable to run our own thermal models for this object.

For the binary 2001 $QC_{298}$, color photometry in the V-I interval was unavailable so we had to use V-J photometry from Stephens *et al.* (2003) to estimate the spectral slope. The object is likely to be somewhat more red in the V-I interval than it is over the broader V-J interval.

For Pluto and Charon, we took diameters from Tholen *et al.* (1987) and Buie *et al.* (1992) to bound possible sizes. Albedos were taken from Buie *et al.* (1997) and Buie and Grundy (2000). Pluto's albedo envelope reflects its rotational lightcurve. Spectral slopes *s* were computed from spectra from Fink and Disanti (1988) and Grundy and Fink (1996).

## 4. Comparisons

Comparing R-band albedo with size *d*, spectral slope *s*, and inclination *i* (with respect to the invariable plane), no clear albedo trends are evident with size or with color, especially if the planet Pluto is excluded (Pluto's high albedo being attributed to surface-atmosphere interactions unlikely to affect the smaller TNOs). Higher albedos among smaller TNOs and among TNOs having higher inclinations and eccentricities could result from higher collisional erosion rates on such objects, if more pristine sub-surface materials are brighter than space-weathered surfaces (e.g., Stern 1995; Durda and Stern 2000; Strazzulla *et al.* 2003). Below some threshold diameter, erosion rates could be expected to exceed photolytic and radiolytic darkening timescales, producing a characteristic signature in a plot of albedo versus size. The two highest albedos in our sample are for objects smaller than 200 km, but there are as yet too few small objects to make a convincing case for what may eventually prove to be distinct albedo distributions between smaller and larger TNOs.

The existence of distinct dynamical classes of TNOs could potentially obscure trends affecting particular sub-populations coming from different source regions or experiencing different thermal, collisional, or radiation environments. Based on their average behavior during 10 Myr orbital integrations, we divided our sample into three dynamical classes to help distinguish possible source populations or environmental influences. Objects found to inhabit mean motion resonances with Neptune were classed as Resonant. Non-resonant objects with Tisserand parameters with respect to Neptune greater than three and mean eccentricities less than 0.2 were classified as Classical, and non-resonant objects with greater mean eccentricities were classified as Scattered (e.g., Chiang *et al.* 2003; Elliot *et al.* 2005). These three classes: Classical, Resonant, and Scattered, are colored red, blue, and green, respectively, in Fig. 1. All three groups are seen to have diverse albedos, and we do not see evidence for systematic albedo differences between Classical objects, with their small inclinations and eccentricities, and Scattered and Resonant objects, which have larger inclinations and eccentricities, and thus higher mean collision speeds.

Note to Editor: Figure 1 should go somewhere near here.

A dark, 4% albedo has historically been assumed for TNOs by analogy to comet nuclei (e.g., Campins and Fernández 2002). While low albedos do exist among our sample, much higher albedos are also seen, with the mean R albedo of our sample being 14%, excluding Pluto and Charon and objects having only lower limits. These higher albedos are consistent with models for impact formation of binary TNOs, which need smaller, brighter objects to produce enough binaries (Stern 2002b).

Distinguishing size effects from effects of duplicity, dynamical class, or other possible influences is a potentially serious difficulty. The seven smallest objects in our sample are all primaries of binary systems. Binary objects could conceivably have unusual albedos resulting from the fragmentation of larger, differentiated objects in binary-forming impact events or from the re-accretion of the debris thrown off during such a violent event. A sample of larger binaries would provide a useful experimental control to help distinguish size effects from the effects of duplicity, but based on the limited data in hand, we see no compelling evidence for differences between the albedos of small binaries and the larger, singular objects probed by thermal observations.

Curiously, the lowest mean inclinations among Classical KBOs in our sample correspond to the highest albedo objects, as seen in red in the bottom panel of Fig. 1, suggesting an anti-correlation between albedo and inclination among the small, Classical KBO sub-population. Such a pattern, if confirmed by additional data, could perhaps relate to well-established correlations between color and inclination (or mean impact speed) among these objects (e.g., Hainaut and Delsanti 2002; Stern 2002a; Trujillo and Brown 2002).

One might expect the albedos of Resonant TNOs (in blue in Fig. 1) to differ from those of the lower inclination Classical objects, based on the broader range of colors among 3:2 Resonant TNOs (e.g., Doressoundiram and Boehnhardt 2003; Fulchignoni and Delsanti 2003) as well as the possibility that Resonant objects formed closer to the sun and were propelled outward by resonance sweeping during the outward migration of Neptune (e.g., Malhotra 1995; Morbidelli and Levison 2003). The five Resonant TNOs in our sample (four in the 3:2 resonance and one in the 2:1 resonance) exhibit diverse albedos, just as the Classical KBOs do, ranging from very dark (15789, R albedo 3.5% ± 1.7%) to rather bright, with the binary (47171, R albedo 22% ± 10%) being one of the brighter Resonant TNOs. From this small sample, the range of R albedos of Resonant TNOs is statistically indistinguishable from that of Classical KBOs, but if there is any trend with inclination, it is the opposite of that suggested by the data for Classical KBOs, with the lowest albedo Resonant TNO also having the lowest inclination, as shown in the bottom panel of Fig. 1. Inclinations of 3:2 Resonant TNOs are thought to be related to the heliocentric distances at which individual objects formed (e.g., Gomes 2000, 2003), but there is little evidence in our data for correlation of albedo with inclination among objects inhabiting this resonance.

All of the Scattered objects except (90377) Sedna have Tisserand parameters below three, indicating that their orbits are unstable with respect to perturbation by Neptune. As a result, their population could have been drawn from multiple sources, potentially obscuring trends detectable among the Classical or Resonant populations. Although the majority of the Scattered objects in our sample only have limits, they appear to exhibit albedo diversity similar to that of other dynamical classes.

Determination of TNO albedos is necessary for conversion of magnitude-frequency to size-frequency distributions. If albedo were eventually found to correlate with object size, it would imply changes in the shape of the size-frequency curve relative to that of the magnitude-frequency curve, with implications for the total mass of the trans-neptunian population (e.g., Bernstein *et al.* 2004). The existence of diverse albedos implies that the lower albedo objects in each absolute magnitude bin represent a disproportionate fraction of the mass in that bin, while the higher albedo objects in a given mass bin have lower absolute magnitudes and thus higher probabilities of being discovered.

TNO albedos could prove highly revealing of their compositional taxonomy, potentially breaking ambiguities between similarly-colored but compositionally distinct classes of objects. Analogies can be drawn to the impact of radiometric albedos on asteroid taxonomy in the 1980s (e.g., Tholen and Barucci 1989). Albedo determinations are also needed to enable quantitative compositional interpretation of infrared spectra of TNOs (e.g., Grundy and Stansberry 2003).

## 5. Conclusion

Statistical study of TNO albedos is limited by the small number of measurements, but useful hints are beginning to emerge. Contrary to expectation, convincing evidence is not seen for dependence of albedo on object size for diameters spanning an order of magnitude. Similarly, no clear correlation of color with albedo is observed. A wide range of albedos is evident among Scattered, Resonant, and Classical objects, as well as among both small and large, and gray and red objects. Models of size-dependent surface processes such as impact erosion and volatile loss being proposed to interpret TNO colors need to accommodate the existence of bright and dark objects of diverse sizes, colors, and dynamical classes. Visible wavelength albedos of most objects are higher than the 4% value once assumed, calling for revision of mass and size estimates for TNO populations based on 4% albedos. Based on this sample (excluding Pluto and Charon and the objects having only lower limits), a better value would be the median R albedo of 10%, or perhaps an R albedo distribution with a mean of 14% and an average deviation from the mean of 10%.

Spitzer Space Telescope observations can be expected to play a valuable role in future TNO size and albedo studies (e.g., Stansberry *et al.* 2004). However, Spitzer's instruments are not sensitive enough to detect the smaller and higher albedo TNOs that binary studies are beginning to reveal, so binaries will continue to offer unique opportunities for physical studies, especially of smaller TNOs. Resolving possible ambiguities between effects of duplicity and size calls for more sensitive thermal capabilities and/or multi-station stellar occultation observations to provide size estimates of small, non-binary TNOs.

## Acknowledgments

This work was made possible by NASA/ESA Hubble Space Telescope programs #9386, #9585, and #9991, support for which was provided by NASA through a grant from the Space Telescope Science Institute, which is operated by the Association of Universities for Research in Astronomy, Inc., under NASA contract NAS5-26555. Two anonymous referees contributed useful input. We are also grateful to M.W. Buie and E.I Chiang for help with dynamical classification, and to the free and open source software communities for providing essential tools used in this work, notably Linux, the GNU tools, MySQL, OpenOffice.org, Tcl/Tk, and FVWM.

## Table 1. TNO Size and Albedo

| Object | Dynamical Class[1] | Spectral Slope | Method | Diameter (km)[2] | R Albedo (%) |
|---|---|---|---|---|---|
| (15789) 1993 SC | 3:2 | 35 ± 9 | Thermal (90 μm) | $398^{+108}_{-171}$ | $3.5^{+1.6}_{-1.3}$ |
| (15874) 1996 $TL_{66}$ | Scattered | 0.7 ± 1.5 | Thermal (90 μm) | ≤ 958 | ≥ 1.8 |
| (19308) 1996 $TO_{66}$ | 3:2 | 1.3 ± 1.3 | Thermal (1.2 mm) | ≤ 902 | ≥ 3.3 |
| (19521) Chaos | Classical | 19 ± 1 | Thermal (1.2 mm) | ≤ 747 | ≥ 5.8 |
| (20000) Varuna | Classical | 18 ± 1 | Thermal (0.9, 1.2 mm) | $936^{+238}_{-324}$ | $3.7^{+1.1}_{-1.4}$ |
| (24835) 1995 $SM_{55}$ | Scattered | 2.2 ± 0.6 | Thermal (1.2 mm) | ≤ 704 | ≥ 6.7 |
| (26308) 1998 $SM_{165}$ | 2:1 | 21 ± 1 | Binary orbit | 238 ± 54 | 14 ± 6 |
| (28978) Ixion | 3:2 | 17 ± 1 | Thermal (1.2 mm) | ≤ 822 | ≥ 14.8 |
| (38628) Huya | Scattered | 16 ± 6 | Thermal (1.2 mm) | ≤ 548 | ≥ 8.4 |
| (47171) 1999 $TC_{36}$ | 3:2 | 23 ± 1 | Binary orbit | 301 ± 68 | 22 ± 10 |
| (50000) Quaoar | Classical | 20 ± 1 | Direct imaging | 1260 ± 190 | 10 ± 3 |
| (55565) 2002 $AW_{197}$ | Scattered | 17 ± 1 | Thermal (1.2 mm) | $886^{+115}_{-131}$ | $10.1^{+3.8}_{-2.2}$ |
| (55636) 2002 $TX_{300}$ | Scattered | −0.5 ± 1.4 | Thermal (1.2 mm) | ≤ 709 | ≥ 19 |
| (58534) 1997 $CQ_{29}$ | Classical | 18 ± 3 | Binary orbit | 77 ± 18 | 39 ± 17 |
| (66652) 1999 $RZ_{253}$ | Classical | 25 ± 3 | Binary orbit | 170 ± 39 | 29 ± 12 |
| (84522) 2002 $TC_{302}$ | Scattered | - | Thermal (1.2 mm) | ≤ 1211 | ≥ 5.1 |
| (88611) 2001 $QT_{297}$ | Classical | 20 ± 2 | Binary orbit | 168 ± 38 | 10 ± 4 |

| Object | Dynamical Class[1] | Spectral Slope | Method | Diameter (km)[2] | R Albedo (%) |
|---|---|---|---|---|---|
| (90377) Sedna | Scattered | 36 ± 2 | Thermal & imaging | ≤ 1800 | ≥ 4.6 |
| 1998 $WW_{31}$ | Classical | 0.5 ± 3.0 | Binary orbit | 152 ± 35 | 6.0 ± 2.6 |
| 2001 $QC_{298}$ | Scattered | 4 ± 1 | Binary orbit | 244 ± 55 | 2.5 ± 1.1 |
| Pluto | 3:2 | 11 ± 2 | Various | 2200 ± 90 | 72 ± 12 |
| Charon | 3:2 | −2 ± 2 | Various | 1260 ± 90 | 37 ± 2 |

**Table 1 notes:**

1. "3:2" and "2:1" indicate objects orbiting in 3:2 and 2:1 mean motion resonances with Neptune. "Classical" indicates non-resonant objects having Tisserand parameters with respect to Neptune greater than 3 and mean eccentricities less than 0.2 over 10 Myr orbital integrations (e.g., Chiang *et al.* 2003; Elliot *et al.* 2005). "Scattered" indicates all other objects.

2. For binary objects, diameter is for the brighter/larger component, assuming the two components have equal albedos.

## Table 2. Data Sources

| **Size and albedo from thermal observations** | | |
|---|---|---|
| **Object** | **Source of thermal flux** | **Sources of reflected photometry** |
| (15789) 1993 SC | Th00 | LJ96, TR97, Da00, JL01, |
| (15874) 1996 $TL_{66}$ | Th00 (upper limit only) | Ba99, Da00, Bo01, JL01 |
| (19308) 1996 $TO_{66}$ | Al04 (upper limit only) | Ba99, Da00, Bo01, GL01, JL01 |
| (19521) Chaos | Al04 (upper limit only) | Ba00, Da00, Bo01, De01, Do02 |
| (20000) Varuna | Je01, Le02 | Do02 |
| (24835) 1995 $SM_{55}$ | Al04 (upper limit only) | Bo01, De01, GL01, Do02 |
| (28978) Ixion | Al04 (upper limit only) | Do02, Bo04 |
| (38628) Huya | Al04 (upper limit only) | Do01, JL01, Bo02 |
| (55565) 2002 $AW_{197}$ | Ma02 (see text) | Fo04 |
| (55636) 2002 $TX_{300}$ | Or04 (upper limit only) | Ba00, Bo01, GL01, JL01 |
| (84522) 2002 $TC_{302}$ | Al04 (upper limit only) | (see text) |
| (90377) Sedna | Br04 (see text) | Br04 (see text) |

| **Size and albedo from binary orbit** | | |
|---|---|---|
| **Object** | **Source of binary orbit** | **Sources of reflected photometry** |
| (26308) 1998 $SM_{165}$ | Ma04 | De01 |
| (47171) 1999 $TC_{36}$ | Ma04 | Bo01, De01, Do01, Do03, Te03 |
| (58534) 1997 $CQ_{29}$ | Ma04, No04b | Ba00, Bo01, GL01, JL01 |
| (66652) 1999 $RZ_{253}$ | No04a | De01, Do01 |
| (88611) 2001 $QT_{297}$ | Os03 | Os03 |
| 1998 $WW_{31}$ | Ve02 | Ve02, St03 |
| 2001 $QC_{298}$ | Ma04 | St03 (see text) |

| **References** | |
|---|---|
| LJ96 | Luu and Jewitt (1996) |
| TR97 | Tegler and Romanishin (1997) |
| Ba99 | Barucci *et al.* (1999) |
| Ba00 | Barucci *et al.* (2000) |
| Da00 | Davies *et al.* (2000) |
| Bo01 | Boehnhardt *et al.* (2001) |
| De01 | Delsanti *et al.* (2001) |
| Do01 | Doressoundiram *et al.* (2001) |
| JL01 | Jewitt and Luu (2001) |
| GL01 | Gil-Hutton and Licandro (2001) |
| Bo02 | Boehnhardt *et al.* (2002) |
| Do02 | Doressoundiram *et al.* (2002) |
| Ma02 | Margot *et al.* (2002) |
| Ve02 | Veillet *et al.* (2002) |
| Do03 | Dotto *et al.* (2003) |
| Os03 | Osip *et al.* (2003) |
| St03 | Stephens *et al.* (2003) |
| Te03 | Tegler *et al.* (2003) |
| Al04 | Altenhoff *et al.* (2004) |
| Bo04 | Boehnhardt *et al.* (2004) |
| Br04 | Brown *et al.* (2004) |
| Fo04 | Fornasier *et al.* (2004) |
| Ma04 | Margot *et al.* (2004) |
| No04a | Noll *et al.* (2004a) |
| No04b | Noll *et al.* (2004b) |
| Or04 | Ortiz *et al.* (2004) |

## Figure Caption.

**Fig. 1.** R albedo is plotted versus estimated diameter $d$ (top panel), spectral slope $s$ (middle panel), and mean inclination $i$ over the past 10 Myr (bottom panel) for 20 TNOs plus the planet Pluto and its satellite Charon. Associated uncertainties are approximated by boxes and error bars, and are dotted for objects having only limits. Colors indicate orbital characteristics: red for Classical objects, blue for objects inhabiting mean motion resonances with Neptune, and green for Scattered objects.

# Figure 1.

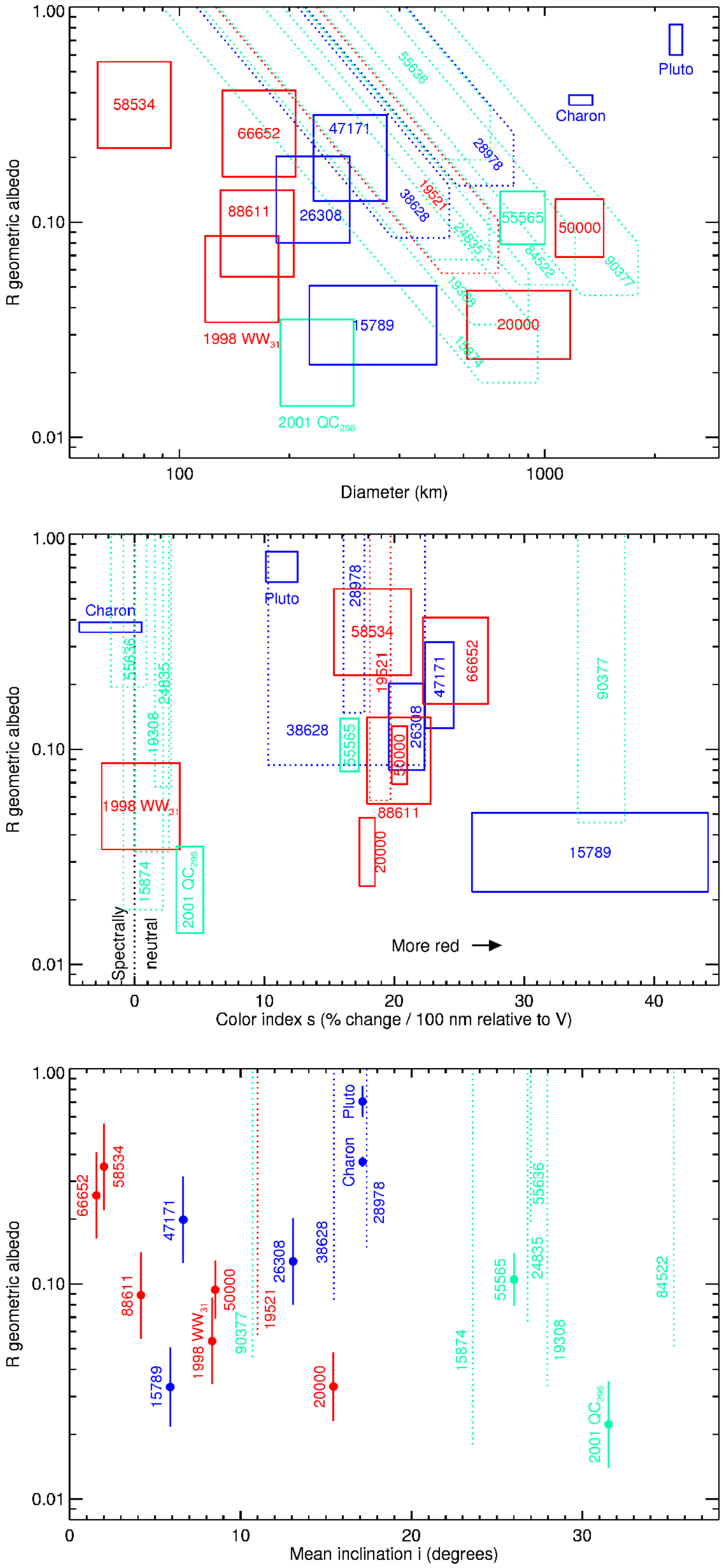
R geometric albedo
1.00
0.10
0.01
100
1000
Diameter (km)
58534
66652
47171
26308
88611
1998 WW31
15789
2001 QC298
20000
50000
55565
Pluto
Charon
55636
28978
19521
38628
24835
84522
90377
19308
15874
0
10
20
30
40
Color index s (% change / 100 nm relative to V)
More red
Spectrally
neutral
Mean inclination i (degrees)